\documentclass[twocolumn,preprintnumbers,superscriptaddress]{revtex4}

\usepackage{amsmath}
\usepackage{dcolumn}
\usepackage{bm}
\usepackage{epsfig}
\usepackage{epstopdf}
\usepackage{graphicx}
\usepackage{amsfonts}
\usepackage{amssymb}
\usepackage{mathrsfs}
\usepackage{color}
\usepackage{multirow}
\usepackage{appendix}
\usepackage[colorlinks,linkcolor=blue]{hyperref}

\begin{document}

\title{Determining temperature and Rabi frequency regarding trapped ions in Doppler cooling: An analytic investigation}
\author{L.-L. Yan}
\email{qingnuanbinghe@126.com}
\affiliation{State Key Laboratory of Magnetic Resonance and Atomic and Molecular Physics, Wuhan Institute of Physics and Mathematics, Chinese Academy of Sciences, Wuhan 430071, China}
\affiliation{School of Physics and Engineering, Zhengzhou University, Zhengzhou 450001, China}
\author{Shi-Lei Su}
\affiliation{School of Physics and Engineering, Zhengzhou University, Zhengzhou 450001, China}
\author{Mang Feng}
\email{mangfeng@wipm.ac.cn}
\affiliation{State Key Laboratory of Magnetic Resonance and Atomic and Molecular Physics, Wuhan Institute of Physics and Mathematics, Chinese Academy of Sciences, Wuhan 430071, China}
\affiliation{Department of Physics, Zhejiang Normal University, Jinhua 321004, China}

\begin{abstract}
Doppler cooling with lasers is essential to ions' trapping and also a preliminary step towards achievement of ultracold ions. Due to lack of effective tools, experimentally monitoring the ions' temperature and the laser-ion coupling is difficult in Doppler cooling. Here we analytically explore the Doppler cooling process of trapped ions, exemplified by $^{40}$Ca$^{+}$, by solving the friction coefficient in the Doppler cooling with respect to a thermal bath, particularly, to a bath with large heating rate. We show four regions for cooling and heating induced by the three-level electromagnetically induced transparency and propose a practical method for measuring the Rabi frequency by the Doppler cooling window. In addition, the final temperature of the laser-cooled ions can be obtained analytically in the case of a weak thermal bath, whereas for the strong thermal bath this requires numerical treatment due to involvement of the Doppler shift and large Lamb-Dicke parameter. Our analytic results would help for understanding many experimental observations, such as configuration phase transition, phonon laser and thermodynamics regarding hot trapped ions.
\end{abstract}
\maketitle

\section{introduction}

Cooling the ions by lasers in the electromagnetic potential is essential to many applications, such as stable confinement of the ions for precision measurements \cite{prl-90-143602,prl-114-223001,RMP-114-223001,pra-98-052507} and manipulation of the ions for quantum information processing\cite{prl-116-080502,pra-95-052319,prl-120-220501}. Doppler cooling is the basic technique employed in ion-trap systems and also the preliminary step to further cooling of the ions down to the ultracold states \cite{pra-64-063407,review-ion,pra-69-043402,pra-96-012519}.

Doppler cooling could be generally understood by two levels, in which the laser with frequency tuned slightly below the two-level resonance frequency could gradually dissipate the kinetic energy of the two-level system by means of Doppler effect \cite{wineland1,hansch}. But practically, Doppler cooling usually applies to multi-level systems, in which dark resonance \cite{d1,d2,d3} occurs in the fluorescence signal and cooling efficiency is related to several factors \cite{e1,e2,e3}. To investigate the cooling mechanism, some numerical efforts have been made based on multi-processes of coherent population transfer \cite{tolazzi}. Moreover, most characteristic quantities of the ions during the Doppler cooling are hard to measure accurately. For example, temperature of the ions, in the absence of elaborate sideband spectroscopy, has to be determined by some special means \cite{tolazzi}. Besides, the laser-ion coupling, i.e., the Rabi frequency, in the range from few tenth to hundreds of phonons, is also unavailable to measure.

\begin{figure}[hbtp]
\centering {\includegraphics[width=6 cm, height=4.5 cm]{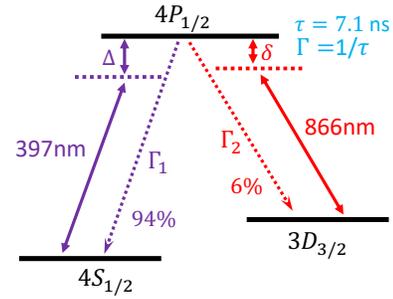}}
\caption{(Color online) Low-lying energy levels of $^{40}$Ca$^{+}$ ion and associated transitions regarding Doppler cooling, where the laser wavelength, the detuning and the branching ratio are labeled in each transition. The lifetime of the excited state is also labeled. The 397 nm laser is called the cooling laser and the 866 nm laser applies for repumping.}
\label{SFig12}
\end{figure}

In this work, we try to propose a practical method to determine the final temperature and the Rabi frequency regarding the ions by an analytic investigation of the Doppler cooling process, which would be of great help for understanding some important characteristics of the ions in hot states. For our purpose of analytic study, we model a three-level system, exemplified by the $^{40}$Ca$^{+}$ ion, whose Doppler cooling consists of two lasers, behaving as an electromagnetically induced transparency (EIT) configuration \cite{prl-85-4458}. In contrast to the multi-level consideration under Zeeman splitting induced by external magnetic field \cite{tolazzi}, our model is analytically solvable and experimentally relevant. We explore the cooling effect on the trapped ions, denoted by friction coefficients and final temperature, in an analytic fashion. Some of the important characteristics of the hot ions, such as the Rabi frequency and temperature variation, can be obtained from our theory. As a result, our analytic results would be very useful for monitoring the trapped ions at temperature in the range from tens of mK to hundreds of mK.

As an example, we sketch energy levels of the trapped $^{40}$Ca$^{+}$ ion system in Fig. \ref{SFig12} that the 397 nm laser with detuning $\Delta$ drives the transition between the excited state $|4P_{1/2}\rangle$ and the ground state $|4S_{1/2}\rangle$, which plays the main role of cooling \cite{SR-6-21547,SA-2-e1600578,prl-120-210601}. The other transition between $|4P_{1/2}\rangle$ and the metastable state $|3D_{3/2}\rangle$, driven by the 866 nm laser with detuning $\delta$, is to repump the 6$\%$ leakage of spontaneous emission back to $|4P_{1/2}\rangle$. The lifetime of $|4P_{1/2}\rangle$ is $\tau=7.1$ ns, corresponding to the decay rate $\Gamma/2\pi=22.4$ MHz. The branching ratios of the decay from $|4P_{1/2}\rangle$ to $|4S_{1/2}\rangle$ and to $|3D_{3/2}\rangle$ are, respectively, $94\%$ and $6\%$, implying $\Gamma_2/\Gamma_1=0.064\ll 1$.

In the following sections, we first analytically discuss the friction coefficient in the Doppler cooling and the lowest temperature reached finally by the Doppler cooling for different baths. To better understand our theory, we start from a general two-level system and derive the universal three-level Doppler cooling formulations in comparison with the two-level counterparts. For the baths with different heating rates, we try below to explore the cooling mechanism analytically in the weak heating case, and numerically for the strong heating situation. Finally, we summarize our theory with a brief discussion.

\section{Solution without thermal bath}

\subsection{Friction Coefficient in a Two-Level System}

\begin{figure}[hbtp]
\centering {\includegraphics[width=8 cm, height=4.5 cm]{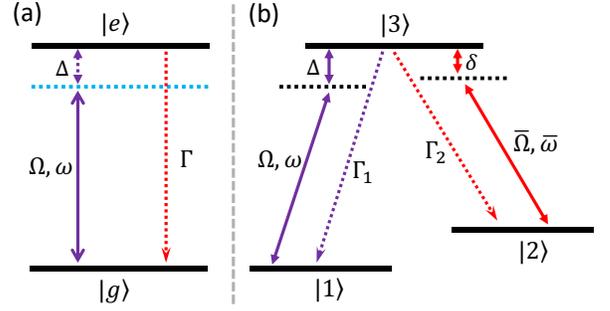}}
\caption{(Color online) (a) Doppler cooling scheme for two levels. A laser with the frequency $\omega$, detuning $\Delta$ and Rabi frequency $\Omega$ drives the transition between the excited state $|e\rangle$ and ground state $|g\rangle$. $\Gamma$ denotes the decay rate from $|e\rangle$ to $|g\rangle$. (b) Doppler cooling scheme for three levels. The two lasers with the detuning $\Delta$ ($\delta$) and Rabi frequency $\Omega$ ($\bar{\Omega}$) drive the transitions between the ground state $|1\rangle$ ($|2\rangle$) and the excited state $|3\rangle$. $\omega$ and $\bar{\omega}$ denote the frequencies of the two lasers, respectively. The decay rates from the excited state $|3\rangle$ to $|1\rangle$ and $|2\rangle$ correspond to $\Gamma_1$ and $\Gamma_2$.  }
\label{SFig1}
\end{figure}

A Doppler cooling process for a two-level system is plotted in Fig. \ref{SFig1}(a). In a rotating frame with respect to the laser frequency, the Hamiltonian for $z$-axis motion is written as ($\hbar=1$),
\begin{equation}
H_{1}^r=-\frac{\Delta}{2}\sigma_z+\frac{\Omega}{2}(|e\rangle\langle g|e^{ik_1z\cos\theta}+h.c.),
\label{Eq1}
\end{equation}
where $\Delta=\omega-\omega_z$ is the detuning of the laser frequency $\omega$ with respect to the two-level resonance frequency $\omega_z$, $\sigma_{z,x}$ are usual Pauli operators for the two levels $|g\rangle$ and $|e\rangle$, $\Omega$ is the laser-induced Rabi frequency and the laser beam in the $xz$ plane has an intersection angle $\theta$ with respect to $z-$axis. $k_{1}$ is the wave number of the cooling laser. The average force in the steady state $\rho_s$ is given by $\langle F\rangle=-\langle \nabla H_{1}^r\rangle$, that is,  \begin{equation}
\langle F\rangle=-\frac{\Omega k_1 \cos\theta}{2}(i\text{Tr}[\rho_s |e\rangle\langle g|] e^{ik_1 z\cos\theta}+h.c.),
\label{Eq3}
\end{equation}
which can be reduced, due to $k_1 z\cos\theta\ll 1$ after the Doppler cooling, to,
\begin{equation}
\langle F\rangle=-\frac{\Omega k_1 \cos\theta}{2}\text{Tr}[\rho_s \sigma_y].
\label{Eq4}
\end{equation}
So the average force $\langle F\rangle$ can be solved if the solution of $\text{Tr}[\rho_s \sigma_y]$ is available (See Appendix A for more details).

\begin{figure}[hbtp]
\centering {\includegraphics[width=8.0 cm, height=5.0 cm]{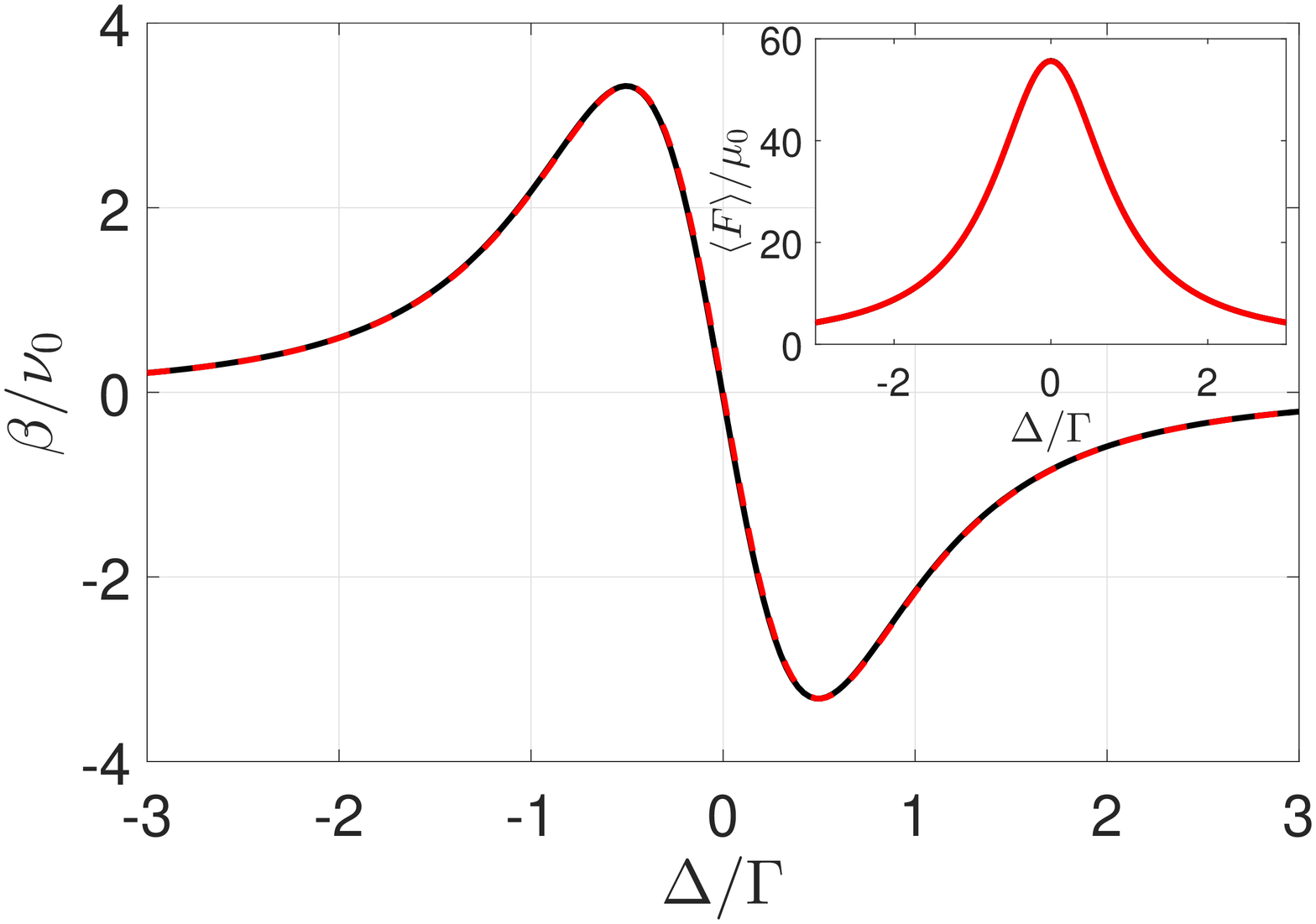}}
\caption{(Color online) Friction coefficient as a function of the detuning $\Delta$. The red dashed and black solid curves denote, respectively, the analytic result (Eq. (\ref{Eq11})) and numerical result (Eq. (\ref{Eq12}) in Appendix B). Inset: the force as a function of the detuning $\Delta$ by numerical solution (black solid curve) and analytic solution (red dashed curve), respectively. The units of force and friction coefficient are $\mu_0=10^{-21} $ N and $\nu_0=10^{-21}$ N$\cdot$ s$/$m.  The parameters are set as $\Gamma/2\pi=22.4$ MHz, $k_1=2\pi/397\times 10^9$ m$^{-1}$, $\theta=\pi/4$, $v_z=0$, and $\Omega=\Gamma$, respectively.}
\label{SFig2}
\end{figure}

Using the saturation parameter $s:=2\Omega^2/\Gamma^2$, Eq. (\ref{Eq4}) can be rewritten as
\begin{equation}
\langle F\rangle=k_1 \cos\theta\frac{\Gamma}{2}\frac{s}{1+s+4(\bar{\Delta}/\Gamma)^2},
\label{Eq10}
\end{equation}
where the velocity-dependent detuning is defined as $\bar{\Delta}= k_1v_z\cos\theta-\Delta$ with the velocity $v_z$ satisfying $k_1v_z\cos\theta\ll\Delta$. If $\bar{\Delta}\rightarrow 0 $ and $ s\rightarrow \infty$, we have $\langle F\rangle=(1/2) \Gamma k_1\cos\theta$, which happens in the case of the maximum momentum change due to absorption of a single photon in unit time. Besides, expanding the force as $\langle F\rangle=F_0-\beta v_z+O(v_z^2)$ at the point $v_z=0$, we obtain the constant force $F_0=k_1\Gamma s\cos\theta/2[1+s+4(\Delta/\Gamma)^2]$ and the friction coefficient
\begin{equation}
\beta=-\frac{4(k_1 \cos\theta)^2s \Delta}{\Gamma(1+s+4\Delta^2/\Gamma^2)^2}.
\label{Eq11}
\end{equation}
From Eq. (\ref{Eq11}), we see that, for a given $s$, the friction coefficient $\beta$ reaches the maximum value at $\Delta=-\sqrt{(1+s)\Gamma^2/12}$, while for a given $\Delta$, $\beta$ is maximized  at $s=1+4\Delta^2/\Gamma^2$. If both of the two above conditions are satisfied, we have $\beta_{\text{max}}=(k_1 \cos\theta)^2/4$ when $\Delta=-\Gamma/2$ and $s=2$ (see Fig. \ref{SFig2}). On the other hand, it can be seen that a red-detuned laser ($\Delta<0$) creates cooling whereas the blue-detuned laser ($\Delta>0$) yields heating, as plotted in Fig. \ref{SFig2}.

\subsection{Friction Coefficient in a Three-Level System }

Extending above treatment to a three-level system, as showed in Fig. \ref{SFig1}(b), we add an auxiliary repumping laser to drive the population between the levels $|2\rangle$ and $|3\rangle$. Under a unitary transformation, i.e., $H_{2}^r=e^{-iRt}He^{iRt}+R$ with $R=\omega |1\rangle\langle 1|+\bar{\omega}|2\rangle\langle 2|$, $H$ turns to be
\begin{eqnarray}
H_{2}^r&=&\Delta |1\rangle\langle 1|+\delta |2\rangle\langle 2|+\frac{\Omega}{2}(|3\rangle\langle 1|e^{ik_1z\cos\theta}+h.c.)+ \notag \\
&&\frac{\bar{\Omega}}{2}(|3\rangle\langle 2|e^{-ik_2z\cos\theta\cos\bar{\theta}}+h.c.),
\label{Eq14}
\end{eqnarray}
where $\bar{\theta}$ is the intersection angle between the repumping laser and the $x$-$z$ plane \cite{note1}, $\Delta=\omega_1+\omega-\omega_3$ and $\delta=\omega_2+\bar{\omega}-\omega_3$ with $\omega_{i}$ the energy of the level $|i\rangle$. $\sigma_x^{ij}=|i\rangle\langle j|+|j\rangle\langle i|$ and $k_{1}$ ($k_{2}$) is the wave number of the cooling (repumping) laser. In the steady state, using $\langle F\rangle=-\langle \nabla H_{2}^r\rangle$, we obtain the average force as
\begin{eqnarray}
\langle F\rangle&=&-\frac{\Omega k_1 \cos\theta}{2}(i\text{Tr}[\rho_s |3\rangle\langle 1|] e^{ik_1 z\cos\theta}+h.c.) + \label{Eq15} \\
&&\frac{\bar{\Omega} k_2 \cos\theta\cos\bar{\theta}}{2}(i\text{Tr}[\rho_s |3\rangle\langle 2|]  e^{-ik_2 z\cos\theta_2\cos\bar{\theta}}+h.c.),  \notag
\end{eqnarray}
where $\rho_s$ is the steady state of the three-level system. In general, we have $k_1 z\cos\theta,k_2 z\cos\theta\cos\bar{\theta} \ll 1$ for the steady state in Doppler cooling limit. Thus, we obtain
\begin{equation}
\langle F\rangle=-\frac{\Omega k_1 \cos\theta}{2}\text{Tr}[\rho_s \sigma_y^{13}] + \frac{\bar{\Omega} k_2 \cos\theta\cos\bar{\theta}}{2}\text{Tr}[\rho_s \sigma_y^{23}],  \label{Eq16}
\end{equation}
where $\sigma_y^{kj}=i|j\rangle\langle k|-i|k\rangle\langle j|$. More specific solution of Eq. (\ref{Eq16}) can be found in Appendix C.

\begin{figure}[hbtp]
\centering {\includegraphics[width=8.5 cm, height=5.5 cm]{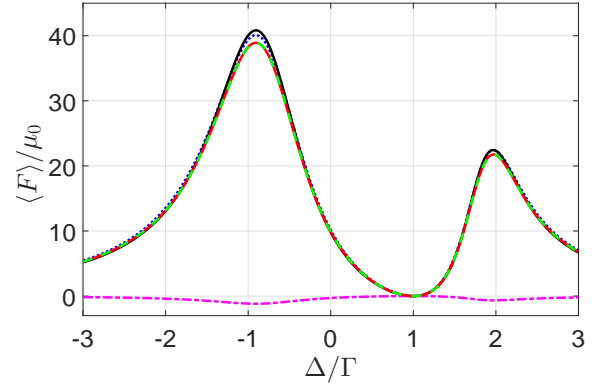}}
\caption{(Color online) Forces as functions of the detuning $\Delta$. The black solid, magenta dotted-dashed, blue dotted, red solid and green dashed curves denote the first term of the force in Eq. (\ref{Eq16}), the second term of the force in Eq. (\ref{Eq16}), the analytic force in Eq. (\ref{Eq23}), the total force in Eq. (\ref{Eq16}) and the numerical calculation obtained by master equation, respectively. The parameters are set as $v_z=0$, $k_1=2\pi/397\times 10^9$ m$^{-1}$, $k_2=2\pi/866\times 10^9$ m$^{-1}$, $\Gamma/2\pi=22.4$ MHz, $\Omega=\Gamma$, $\bar{\Omega}/\Omega=2.5$, $\delta=\Omega$, $\Gamma_1=0.94\Gamma$ and $\Gamma_2=0.06\Gamma$, respectively.}
\label{SFig4}
\end{figure}

In the weak dissipation limit of $\Gamma_2/\Gamma_1\rightarrow 0$, $\text{Tr}[\rho_s \sigma_y^{13}]\approx 0$ and $\langle F\rangle$ can be obtained by inserting the solution of $\text{Tr}[\rho_s \sigma_y^{13}]$ (from Eq. (\ref{Eq22}) in Appendix C) into Eq. (\ref{Eq16}), that is,
\begin{equation}
\langle F\rangle=\frac{4(\bar{\Delta}-\bar{\delta})^2\Omega^2\Gamma_1 k_1 \cos\theta}{\bar{N}},
\label{Eq23}
\end{equation}
where the denominator is $\bar{N}=4(\bar{\Delta}-\bar{\delta})^2(4\bar{\Delta}^2+2\Omega^2+\Gamma_1^2)+8\bar{\Delta}\bar{\Omega}^2(\bar{\delta}-\bar{\Delta})+(\Omega^2+\bar{\Omega}^2)^2$ with the velocity-dependent detuning $\bar{\Delta}=\Delta-k_1v_z\cos\theta$ and $\bar{\delta}=\delta+k_2v_z\cos\theta\cos\bar{\theta}$ satisfying $ \Delta\gg k_1v_z\cos\theta,k_2v_z\cos\theta\cos\bar{\theta} $. In Fig. \ref{SFig4}, the above analytic results are verified numerically by master equation. We find that the contribution from the repumping laser is very small, since the first term of Eq. (\ref{Eq16}) (black solid curve) gives the result very close to the exact solution. In addition, the analytic solution (blue dotted curve) in Eq. (\ref{Eq23}) obtained in the limit $\Gamma_2/\Gamma_1\rightarrow 0$ also agrees with the exact solution very well.

If we expand the force in Eq. (\ref{Eq23}) around the point $v_z=0$, i.e., $\langle F\rangle=F_0-\beta v_z+O(v_z^2)$, a constant force reads $F_0=4(\Delta-\delta)^2\Omega^2\Gamma_1 k_1 \cos\theta/N$ with $N=4(\Delta-\delta)^2(4\Delta^2+2\Omega^2+\Gamma_1^2)+8(\delta-\Delta)\Delta\bar{\Omega}^2+(\Omega^2+\bar{\Omega}^2)^2$. In this case, the friction coefficient turns to be,
\begin{equation}
\beta=4\Omega^2\Gamma_1k_1\cos^2\theta(\frac{N_2}{N}-\frac{N_1}{N^2}),
\label{Eq24}
\end{equation}
where $N_1=8(\Delta-\delta)^2[\bar{\Omega}^2(\delta k_1-2\Delta k_1-\Delta k_2\cos\bar{\theta})+4\Delta(\Delta-\delta)^2 k_1+(\Delta-\delta)(k_1+k_2\cos\bar{\theta})(4\Delta^2+2\Omega^2+\Gamma_1^2)]$ and $N_2=2(\Delta-\delta)(k_1+k_2\cos\bar{\theta})$. For the case of $\delta=\Delta$, we have $\beta=0$, which is the dark resonance observed previously in various experiments, implying that the laser has no cooling effect \cite{tolazzi}. Moreover, we have to mention that a more exact analytic solution for the friction coefficient can be obtained by substituting Eq. (\ref{Eq20}) into Eq. (\ref{Eq16}) and expanding $\langle F\rangle$ to the first order of $v_z$. Such a solution can fit the numerical results better, but with a more complicated form than Eq. (\ref{Eq24}).

\begin{figure}[hbtp]
\centering {\includegraphics[width=8.5 cm, height=7.5 cm]{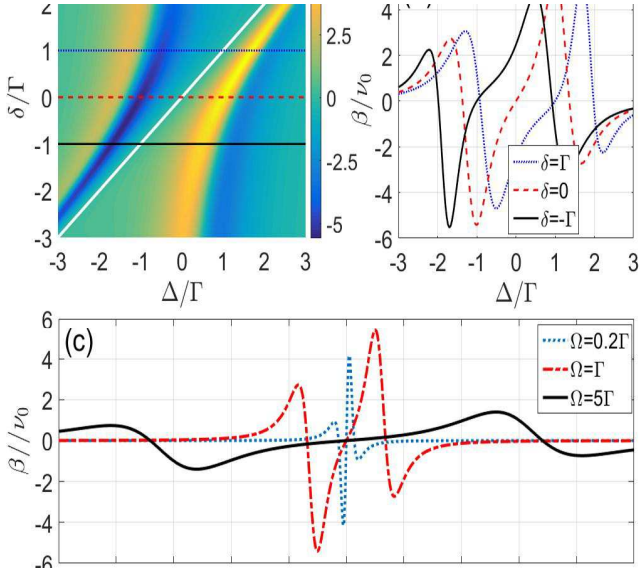}}
\caption{(Color online) (a) Friction coefficient as functions of the detunings $\Delta$ and $\delta$, where the white line denotes the boundary corresponding to the EIT point ($\Delta=\delta$). (b) Three special cases with $\delta=-\Gamma,0,\Gamma$ in (a). (c) Friction coefficient as a function of the detuning $\Delta$ for different values of Rabi frequency $\Omega$ with $\delta=0$. Other parameters are the same as in Fig. \ref{SFig4}. Positive (negative) values of the friction coefficient mean cooling (heating) effect. }
\label{SFig5}
\end{figure}

For clarity, we plot the phase diagram for the friction vs the detunings $\Delta$ and $\delta$ in Fig. \ref{SFig5}(a), where the diagonal line corresponds to $\Delta=\delta$, implying $\beta=0$ due to the EIT that no photon absorption occurs. Besides, for a given detuning $\delta$, the change of $\Delta$ can also yield $\beta=0$ by satisfying the condition $N_1-NN_2=0$ (see Eq. (\ref{Eq24})), which gives two solutions mathematically on either side of the EIT point, i.e., $\Delta=\delta$. The detuning $\Delta$ across these three solutions of $\beta=0$, as plotted in Fig. \ref{SFig5}(b), switches repeatedly the sign of $\beta$ which means alternate occurrence of heating ($\beta<0$) and cooling ($\beta>0$). The latter two solutions are related to the Rabi frequency $\Omega$ (see Fig. \ref{SFig5}(c)), and a suitable Rabi frequency $\Omega$ ($\approx \Gamma$) creates a large friction coefficient.

\subsection{Final Temperature}

\begin{figure}[hbtp]
\centering {\includegraphics[width=8.5 cm, height=7.0 cm]{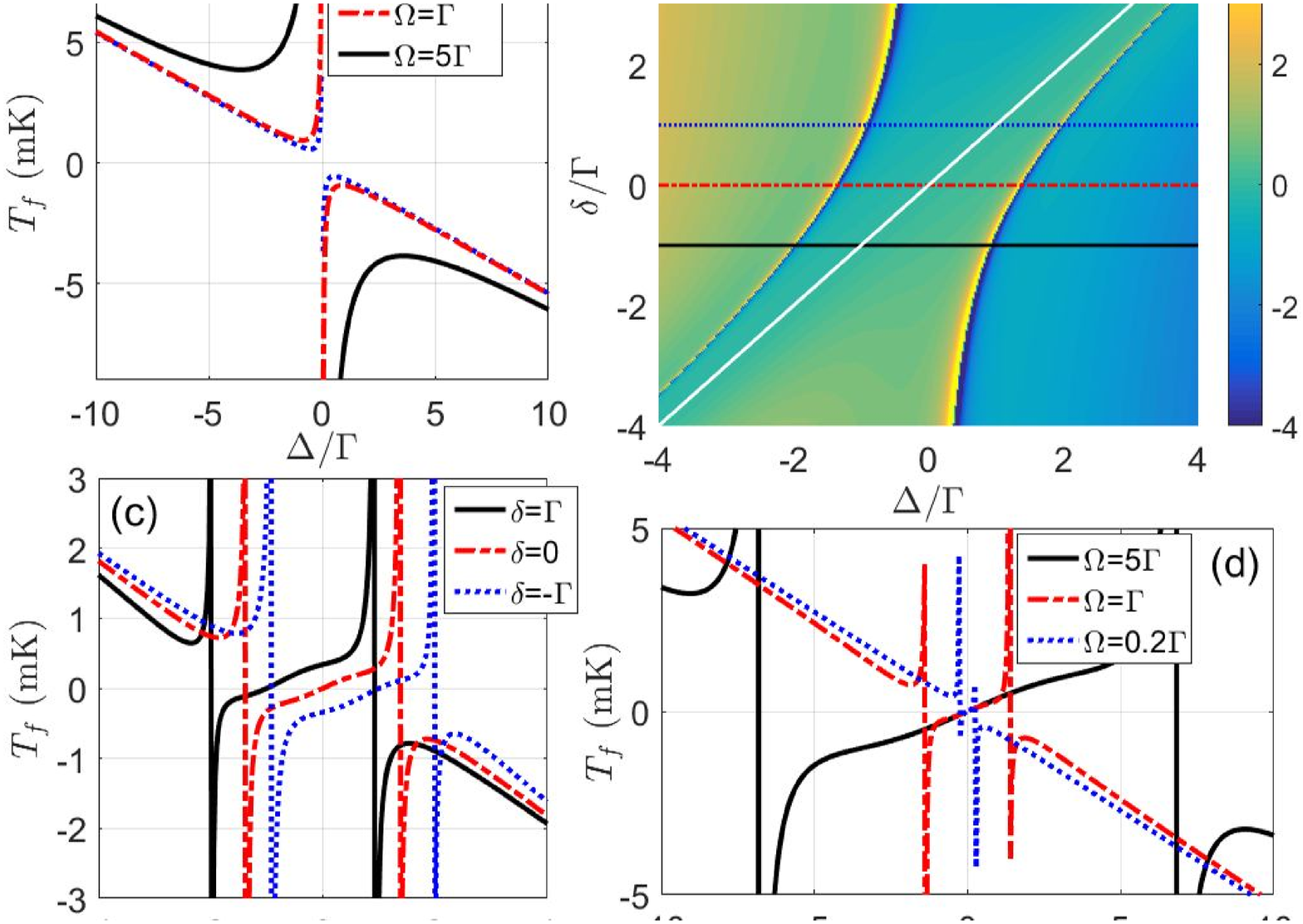}}
\caption{(Color online) (a) Final temperature as a function of the detuning $\Delta$ for two-level system. (b) Final temperature as functions of the detunings $\Delta$ and $\delta$ for three-level system, where the white line denotes the transparency boundary $\Delta=\delta$. (c) Three special cases: $\delta=-\Gamma,0,\Gamma$ in (b). (d) Final temperature in a three-level system as a function of the detuning $\Delta$ for different values of Rabi frequency $\Omega$ with $\delta=0$. Other parameters are the same as in Fig. \ref{SFig4}. The negative temperature, originated from $\beta<$0, implies heating.}
\label{SFig6}
\end{figure}

Spontaneous emission of the trapped ions induces the diffusion of wave packet which leads to a heating effect. The diffusion coefficient $D$ for the two-level system is written as
\begin{equation}
D_2=\frac{k_1^2\cos^2\theta\Omega|\langle\sigma_y\rangle|}{4m^2}
\label{Eqs11}
\end{equation}
and for the three-level system is
\begin{equation}
D=\frac{k_1^2\cos^2\theta\Omega|\langle\sigma_y^{13}\rangle|+k_2^2\cos^2\theta\cos^2\bar{\theta}\bar{\Omega}|\langle\sigma_y^{23}\rangle|}{4m^2},
\label{Eqs12}
\end{equation}
where the values of $\langle\sigma_y\rangle$, $\langle\sigma_y^{13}\rangle$ and $\langle\sigma_y^{23}\rangle$ are available at $v_z=0$. In the long time limit, the diffusion of velocity in time interval $\tau$ is
\begin{equation}
(\Delta v)^2=2D\tau.
\end{equation}
On the other hand, the friction coefficient leads to the velocity decrease which is given by
\begin{equation}
\Delta v=-\beta v_{\infty}\tau/m.
\end{equation}
Towards the end of the cooling process, these two velocities should be of the same order of magnitude, that is, $\Delta v=v_{\infty}$. Thus, using above two equations, we obtain $v_{\infty}=\sqrt{2mD/\beta}$. Besides, the final temperature is determined by $T_fk_B/2=mv_{\infty}^2/2$, which means
\begin{equation}
T_f=\frac{2m^2D}{k_B\beta}.
\label{Eqs15}
\end{equation}
So for the two-level system, we obtain the final temperature as
\begin{equation}
T_{f}=-\frac{\Gamma^2+2\Omega^2+4\Delta^2}{8k_B\Delta}.
\label{Eq25}
\end{equation}
The maximum friction coefficient occurred at $\Omega=\Gamma$ and $\Delta=-\Gamma/2$ (see Fig. \ref{SFig6}(a)) corresponds to the temperature $T_f=\Gamma/k_B=1.07$ mK in the case of $\Gamma/2\pi=22.4$ MHz, and the final energy at this point is $\varepsilon^{\text{min}}_f=mv_{\infty}^2/2=\Gamma$ \cite{RMP-58-699}. Moreover, for a given $\Omega$, the temperature $T_f$ has the smallest value $\sqrt{\Gamma^2+2\Omega^2}/2k_B$ at $\Delta=\sqrt{\Gamma^2+2\Omega^2}/2$ which means $T_f=\sqrt{3}\Gamma/2k_B$ for $\Omega=\Gamma$. In the weak saturation limit $s\rightarrow 0$, i.e., $\Omega/\Gamma\rightarrow 0$, the final temperature is $T_f=\Gamma/2k_B$. On the other hand, the blue-detuned laser ($\Delta>0$) produces a negative friction coefficient $\beta$. Combined with the positive diffusion coefficient, $D$ leads to a negative value of temperature $T_f$, implying the heating effect. In this context, the ions would escape from the trap when $T_f <$0. As a result, the negative values of $T_f$ are not real temperature to be measured.

For the three-level system, under the condition of $\Gamma_1\gg\Gamma_2$, we employ $D=k_1^2\cos^2\theta\Omega|\langle\sigma_y^{13}\rangle|/4m^2$ and the friction coefficient $\beta$ given in Eq. (\ref{Eq24}). Therefore, we obtain
\begin{equation}
T_f=\frac{(\Delta-\delta)^2}{k_BN_h},
\label{Eq26}
\end{equation}
where $N_h=N_2/k_1\cos\theta-N_1/Nk_1\cos\theta$. In Fig. \ref{SFig6}(b), we plot the phase diagram of temperature in variation with the detunings $\Delta$ and $\delta$ under the condition of $\Omega/\bar{\Omega}=0.4$. Around the EIT point for $\beta=0$, both $D$ and $\beta$ are very small due to weak absorption of photons, and $D$ is much smaller than $\beta$, i.e., $D/\beta\rightarrow 0$, which leads to a much lower temperature, even close to zero temperature (see Fig. \ref{SFig6}(b,d)). However, at other two zero points of $\beta$, the photons are absorbed and scattered, which creates a nonzero diffusion coefficient $D$. As indicated in Fig. \ref{SFig6}(c,d)), the system is intensively heated around those two points. The negative value of temperature in Fig. \ref{SFig6} corresponds to the negative $\beta$, implying a heating effect. What's more, Fig. \ref{SFig6} shows that there exists a lower final temperature in three-level case than the counterpart of two-level case for an appropriately large red-detuning region (in the left-hand side of the left zero point of $\beta$), where the final temperature varies similarly to the two-level situation in the red-detuning region (see Fig. \ref{SFig6}(a,c)), for example, for $\Omega=\Gamma$, the lowest temperature obtained by Eq. (\ref{Eq25}) is $0.47$ mK, whereas by Eq. (\ref{Eq26}), it is $0.34$ mK.

\subsection{Measuring Rabi Frequency by the Width of Detuning Window}

\begin{figure}[hbtp]
\centering {\includegraphics[width=8.5 cm, height=6.5 cm]{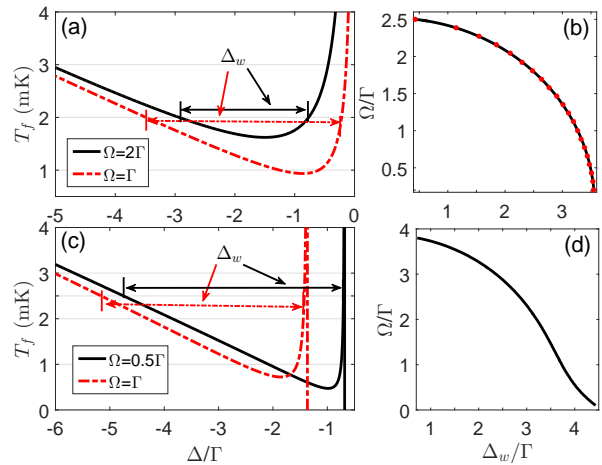}}
\caption{(Color online) (a) and (c) Final temperature as a function of the detuning $\Delta$ for two-level system and three-level system (with $\delta=0$ in three-level system), respectively. The width of detuning window is defined as $\Delta_w=\Delta_{+}-\Delta_-$ where $\Delta_{+} $ and $\Delta_-$ are determined by the two points of the curve at which temperature has a same value, such as $2$ mK in (a) and 2.5 mK in (c). (b) and (d) Rabi frequency as a function of the width of detuning window $\Delta_w$ with a given temperature $2$ mK in (b) and $2.5$ mK in (d), respectively. The dots and solid curves in (b) are, respectively, the numerical calculation and the analytic result in Eq. (\ref{Eq27}), and the solid curve in (d) is the numerical calculation. The parameters are same as in Fig. \ref{SFig4}. }
\label{SFig7}
\end{figure}

As mentioned in the Introduction, the coupling between the cooling laser and the ions in the Doppler cooling process is hard to measure experimentally due to the large linewidth of the excited state. Here we provide a practical method to estimate this coupling, i.e., the Rabi frequency regarding the cooling laser, by measuring the width of the detuning window. Specifically, we first consider the two-level case. Given a temperature $T_f$, we obtain, by utilizing Eq. (\ref{Eq25}), the detuning window $\Delta_w\equiv\Delta_{+}-\Delta_-=\sqrt{4k_B^2T_f^2-(2\Omega^2+\Gamma^2)}$ (also see Fig. \ref{SFig7}(a) for the explanation of the detuning window), which turns to be, \begin{equation}
\Omega=\sqrt{\frac{4k_B^2T_f^2-\hbar^2\Delta_w^2-\hbar^2\Gamma^2}{2\hbar^2}}.
\label{Eq27}
\end{equation}
For a large $T_f$, we have $\Delta_w \gg \Gamma$, indicating $\Omega\propto \Delta_w$, as plotted in Fig. \ref{SFig7}(b). This method can be straightforwardly extended to the three-level case using Eq. (\ref{Eq26}). Considering finite values of $\Gamma_{2}$, we plot Fig. \ref{SFig7}(c,d) numerically.

\section{Solution with thermal bath}

\subsection{Weak Heating Case}

\begin{figure}[hbtp]
\centering {\includegraphics[width=9.2 cm, height=6.0 cm]{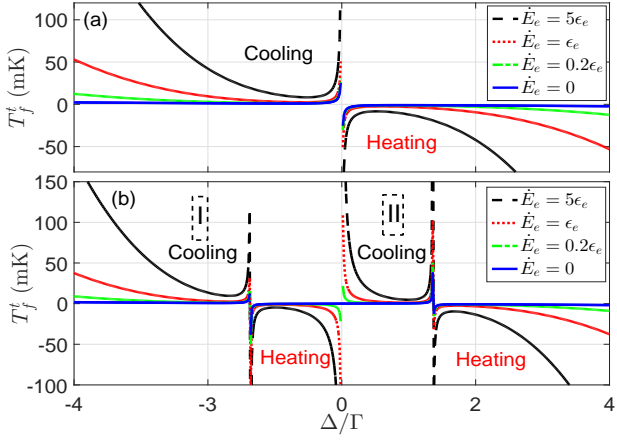}}
\caption{(Color online)  Final temperature as a function of the detuning $\Delta$ under different bath heating effects $\dot{E}_e$ for two-level system (a) and three-level system (b). Parameters are set as $\Omega=\Gamma$ and $\delta=0$. Here the unit of environmental heating effect $\dot{E}_e$ is $\epsilon_e=10^{-21}$N$^2$s$/$kg and the mass of $^{40}$Ca$^+$ion is $m=6.68\times 10^{-26}$ kg. Other parameters are the same as in Fig. \ref{SFig4}. The negative temperature implies heating. }
\label{SFig8}
\end{figure}

The real temperature reachable experimentally is always higher than that predicted in the Doppler cooling limit due to involvement of the unexpected bath noises, such as the fluctuation of electric and magnetic fields, the radio-frequency heating and the collision with background atoms. Here, we assume that the bath is at a high temperature which creates a constant heating effect $\dot{E}_e$. Considering a weak heating, we may simply write the heating effect as $\dot{E}_h=\dot{E}_l+\dot{E}_e$ with the laser scattering effect $\dot{{E}_l}=2mD$. On the other hand, the cooling effect induced by the laser is $\dot{E}_c=-\beta v_{\infty}^2$. In the long time limit, $\dot{E}_c+\dot{E}_h=0$. So we have the final temperature as
\begin{equation}
T^t_f=T_f+\frac{m\dot{E}_e}{k_B\beta}.
\label{Eq28}
\end{equation}
Fig. \ref{SFig8} presents the influence on the ion from the bath noises. As expected, when the heating becomes stronger, the final temperature turns to be higher, especially in the large detuning situation where the friction coefficient is relatively small whereas the heating effect $\dot{E}_e$ keeps constant so that $T_f^t\propto 1/\beta$. On the other hand, at the EIT point $\Delta=\delta$, there is no photon absorption, implying $\beta=0$ and $D=0$. But due to involvement of the bath noise, here we always have $\dot{E}_e>0$, which leads to a high temperature.
We plot both the two-level and three-level cases in Fig. 8. In contrast to the simple curves in two-level case (Fig. 8(a)), three-level case behaves complicated, for example, existence of
a cooling region at the blue detuning (Type II in Fig. \ref{SFig8}(b)), where the temperature has two abrupt changes and also reaches a lower temperature than the first cooling region (Type I in Fig. \ref{SFig8}(b)). In actual experiments, due to the requirement of experimental conditions, the Doppler cooling is always implemented in the first cooling region.

\begin{figure}[hbtp]
\centering {\includegraphics[width=8.5 cm, height=4.5 cm]{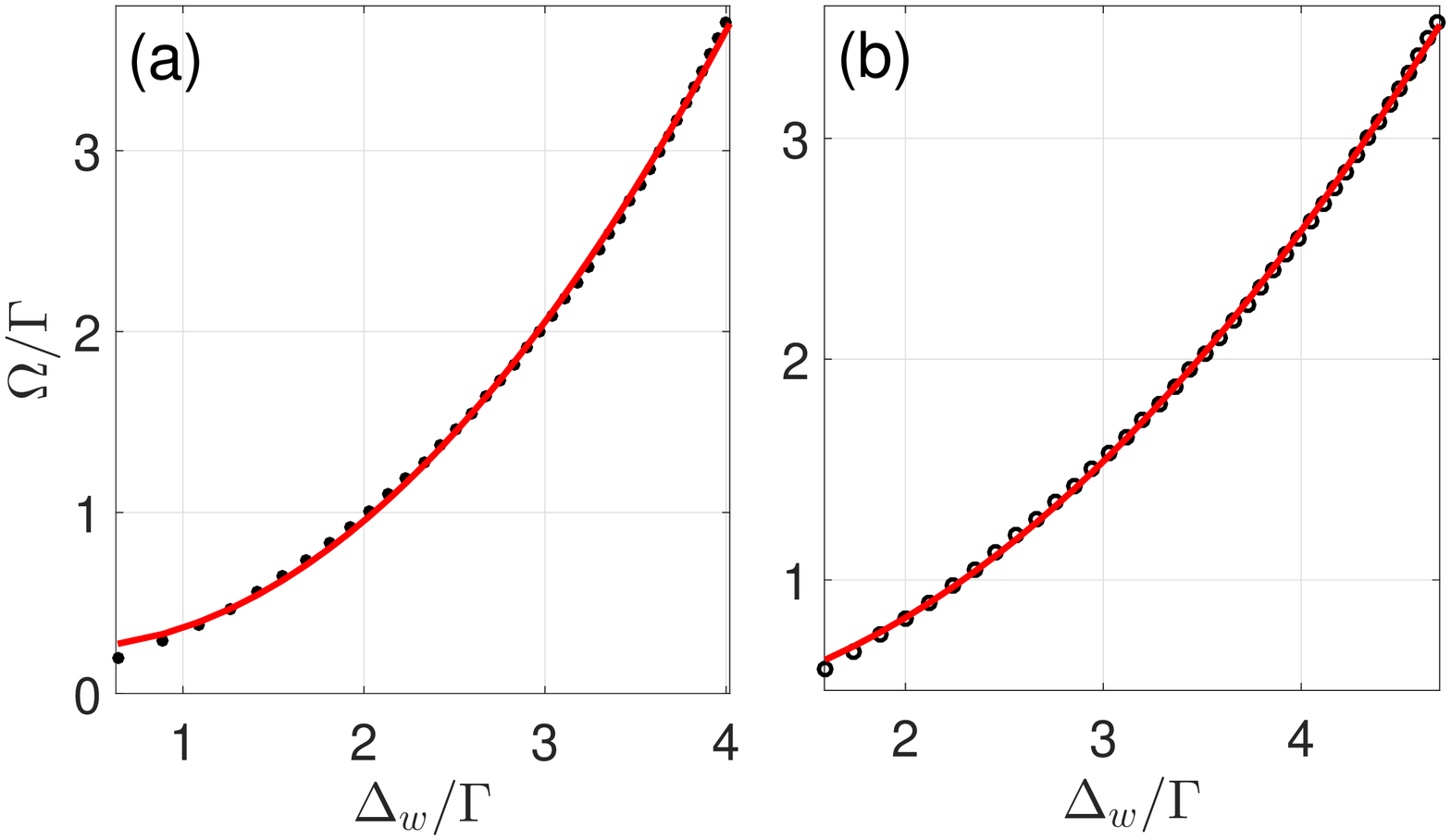}}
\caption{(Color online)  (a) and (b) Rabi frequency as a function of the width of detuning window $\Delta_w$ with a given temperature ($60$ mK in (a) and $100$ mK in (b)) for two- and three-level systems, respectively, under the influence of environmental heating effect. Dots are the numerical calculation and curves denote a fitting by quadratic function. Here $\dot{E}_e=5\times 10^{-21}$ N$^2$s$/$kg and $m=6.68\times 10^{-26}$ kg. Other parameters are same as in Fig. \ref{SFig4}. }
\label{SFig8s}
\end{figure}

On the other hand, in the presence of the environmental heating, the final temperature is mainly determined by the environment, i.e., the second term being dominant in Eq. (\ref{Eq28}). Thus, for a given temperature $T_f$, the detuning window is determined by
\begin{equation}
k_B\beta T_f-m\dot{E}_e=0.
\label{Eq28s}
\end{equation}
Numerical results of Eq. (\ref{Eq28s}) shown in Fig. \ref{SFig8s} demonstrate a quadratic behaviour between the width of detuning window and the Rabi frequency of the laser, which is very different from the situation in absence of environmental heating effect.

\subsection{Strong heating case}

In the strong heating case, we have to consider the thermal effect regarding the velocity of the ions and the detuning. First, we define the velocity-dependent detunings $\Delta_0=(\Delta-k_1v_{T}\cos\theta)$ and $\delta_0=(\delta+k_2v_T\cos\theta\cos\bar{\theta})$, and the position $z=z_0+\delta_z :=v_T t+d v_z t$, where the velocity $v_T$ is temperature-dependent  obeying the Maxwell-Boltzmann distribution of velocity vector, i.e., $p(v_T)=e^{-v_T^2/v_p^2}/(v_p\sqrt{\pi})$ with the most probable velocity $v_p=\sqrt{2k_B T/m}$. The fluctuation velocity $d v_z$ (estimated by $\hbar k_1\cos\theta/m$) and the displacement $\delta_z=d v_z/\Gamma$ stem from the photon absorption. The first part of $\Delta_0$ or $\delta_0$ comes from the ions' temperature which produces a Doppler frequency shift, and the second part contains the momentum effect of the photon which creates the cooling effect and the friction coefficient.

In the present case, the Hamiltonian in Eq. (\ref{Eq14}) is rewritten as
\begin{eqnarray}
H_s^r&=& \Delta_0 |1\rangle\langle 1|+\delta_0 |2\rangle\langle 2|+\frac{\Omega}{2}(|3\rangle\langle 1|e^{ik_1\delta_z\cos\theta}+h.c.) \notag \\
&&+\frac{\bar{\Omega}}{2}(|3\rangle\langle 2|e^{-ik_2\delta_z\cos\theta\cos\bar{\theta}}+h.c.).
\label{Eqs1}
\end{eqnarray}
In general, $z_0\gg\delta_z$ and $v_z^0\gg d v_z$. In the limit of $k_1\delta_z\cos\theta,k_2\delta_z\cos\theta\cos\bar{\theta}\ll 1$, the average force $\langle F\rangle=-\langle \partial H_s^r/\partial\delta_z\rangle$ is given by
\begin{equation}
\langle F\rangle=-\frac{\Omega k_1 \cos\theta}{2}\text{Tr}[\rho_s \sigma_y^{13}] + \frac{\bar{\Omega} k_2 \cos\theta\cos\bar{\theta}}{2}\text{Tr}[\rho_s \sigma_y^{23}], \label{Eqs2}
\end{equation}
which is of the same form as Eq. (\ref{Eq16}).

\begin{figure}[hbtp]
\centering {\includegraphics[width=8.6 cm, height=8.6 cm]{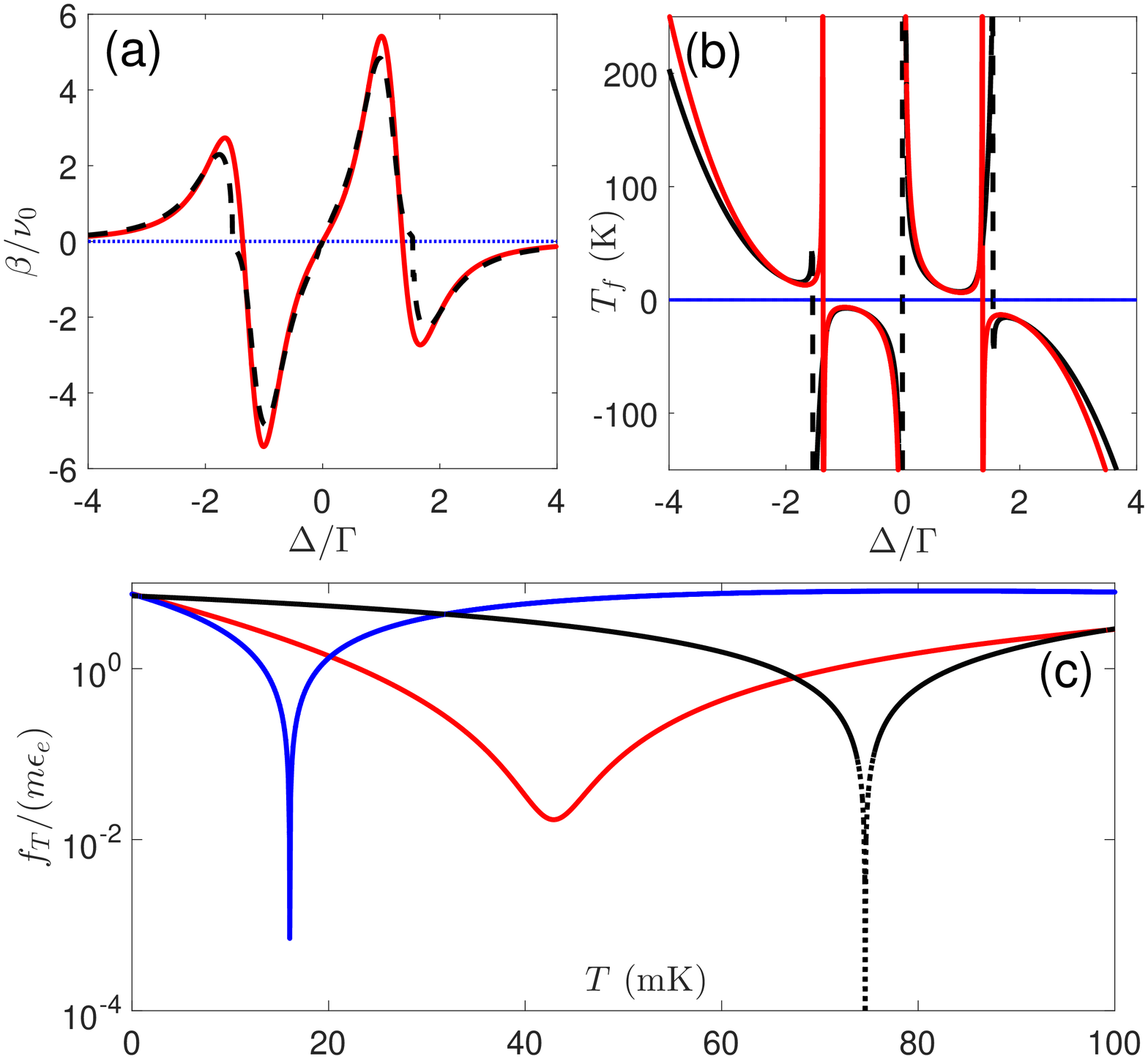}}
\caption{(Color online) Friction $\beta$ (a) and final temperature $T_f$ (b) as functions of the detuning $\Delta$ with $\dot{E}_e=7\times 10^{-21}$ and $m=6.68\times 10^{-26}$ kg. Red and black curves are calculated by Eq. (\ref{Eq28}) and Eq. (\ref{Eqs4}), respectively. (c) Numerical solutions of Eq. (\ref{Eqs4}) for $\Delta/\Gamma=-3$ (black curve), $-1.7 $ (blue curve) and $-1.56$ (red curve) in (a) and (b) with $f_{T}= k_B T\bar{\beta}-m(2m\bar{D}+\dot{E}_e)$. Other parameters are the same as in Fig. \ref{SFig4}.}
\label{SFigs1}
\end{figure}

With the similar process as listed in Appendix C and in the limit $\Gamma_2/\Gamma_1\rightarrow 0$, Eq. (\ref{Eqs2}) is reduced to the same form as Eq. (\ref{Eq23}) only by replacing $\bar{\Delta}$ and $\bar{\delta}$ by $\bar{\Delta}=\Delta-k_1(v_T+d v_z)\cos\theta$ and $\bar{\delta}=\delta+k_2(v_T+d v_z)\cos\theta\cos\bar{\theta}$, respectively. Then expanding the average force around  $dv_z=0$, we have $\langle F\rangle=F_0-\bar{\beta}_{v_T}dv_z+O((dv_z)^2)$ with $F_0$ being a constant of $\langle F\rangle$ at $dv_z=0$, and the friction coefficient
\begin{equation}
\bar{\beta}_{v_T}=4\Omega^2\Gamma_1k_1\cos^2\theta(\frac{\bar{N}_2}{\bar{N}_0}-\frac{\bar{N}_1}{\bar{N}_0^2}),
\label{Eqs3}
\end{equation}
where $\bar{N}_0$ is obtained by replacing $v_z$ in $\bar{N}$ of Eq. (\ref{Eq23}) by $v_T$, $\bar{N}_1$ and $\bar{N}_2$ are obtained from $N_1$ and $N_2$ in Eq. (\ref{Eq24}) by substituting $\Delta$ and $\delta$ by $\Delta_0$ and $\delta_0$, respectively. Compared with $\beta$ in Eq. (\ref{Eq24}), Eq. (\ref{Eqs3}) contains the additional Doppler shift due to thermal effect. Thus, the average friction coefficient at temperature $T$ can be obtained as
\begin{equation}
\bar{\beta}=\int_{-\infty}^{\infty} p(v_T)\bar{\beta}_{v_T}dv_T.
\end{equation}
Combining Eq. (\ref{Eqs15}) and Eq. (\ref{Eq28}), we obtain the final temperature determined by the following equation,
\begin{equation}
k_B T\bar{\beta}-m(2m\bar{D}+\dot{E}_e)=0,
\label{Eqs4}
\end{equation}
where $\bar{D}=\int_{-\infty}^{\infty} p(v_T)\bar{D}_{v_T}dv_T$ with $ \bar{D}_{v_T}$ calculated by Eq. (\ref{Eqs12}) in which $\Delta$ and $\delta$ are replaced by $\Delta_0$ and $\delta_0$, respectively. Specific calculation of this final temperature shows a small red frequency shift with respect to Eq. (\ref{Eq28}) for negative detuning, while a blue frequency shift for positive detuning, see Fig. \ref{SFigs1}.

\section{conclusion}

In simulation of trapped ions' motion, the Langevin equation has usually been employed to describe the stochastic dynamics in the Doppler cooling process, where the dynamical process could be used to study various physical phenomena, e.g., phase transitions of ion-crystal structure \cite{NC-4-2290,NC-4-2291}. In these cases, determining the friction coefficient and finial temperature in the Doppler cooling limit is quite important for understanding the characteristic of the system, particularly due to the fact that these two quantities are potential-independent and thus need to be determined prior to the numerical simulation (see Appendix D).

To summarize, we have developed an analytic investigation for Doppler cooling process of trapped ions subject to thermal noises. Our theory could be immediately applied to ion trap experiments, to determine some important quantities unavailable for detection and to help understanding some experimentally observations regarding hot trapped ions. Moreover, the analytic formulae of the ions' temperature and the Rabi frequency, as presented in our theory, would
help us for discovering the rich and complicated physics at atomic level. Therefore, we consider that our theory developed here would be very useful in hot-ions' experiments for, such as configuration phase transitions, phonon lasers and thermodynamics.

\section*{Acknowledgements}

This work was supported by National Key R$\&$D Program of China under grant No. 2017YFA0304503, by National Natural Science Foundation of China under Grant Nos. 11835011, 11804375, 11734018 and 11674360, and by the Strategic Priority Research Program of the Chinese Academy of Sciences under Grant No. XDB21010100.  \\

\appendix

\section{solution of Tr$[\rho_s\sigma_y]$ for a two-level system}
In order to solve the Bloch equation for the two-level system, we rewrite $H_{1}^r$ in Eq. (\ref{Eq1}) as
\begin{equation}
H^r=\frac{\bar{\Delta}}{2}\sigma_z+\frac{\Omega}{2}\sigma_x,
\label{Eq7}
\end{equation}
where the velocity-dependent detuning is $\bar{\Delta}=(-\Delta+k_1v_z\cos\theta_1)$ with the velocity defined as $v_z=z/t$. Thus, by using the Bloch equation we obtain
\begin{eqnarray}
\frac{d}{dt}\langle\rho_{gg}\rangle&=&\frac{\Omega}{2}\langle\sigma_y\rangle+\Gamma\langle\rho_{ee}\rangle, \notag \\
\frac{d}{dt}\langle\sigma_x\rangle&=&\bar{\Delta}\langle\sigma_y\rangle-\frac{\Gamma}{2}\langle\sigma_x\rangle, \label{Eq8} \\
\frac{d}{dt}\langle\sigma_y\rangle&=&-\bar{\Delta}\langle\sigma_x\rangle+\Omega\langle\sigma_z\rangle-\frac{\Gamma}{2}\langle\sigma_y\rangle, \notag
\end{eqnarray}
where $\rho_{ee}=|e\rangle\langle e|$ and $\rho_{gg}=|g\rangle\langle g|$. The total probability in the diagonal entries satisfies $\rho_{gg}+\rho_{ee}=1$. Solving Eq. (\ref{Eq8}), we obtain
\begin{equation}
\langle\sigma_y\rangle=-\frac{2\Omega\Gamma}{4\bar{\Delta}^2+2\Omega^2+\Gamma^2}.
\label{Eq9}
\end{equation}

\section{the dynamical description of the system }
With the Bloch equation, the dynamics of the system can be described as
\begin{equation}
\frac{d\langle \mathcal{A}(t)\rangle }{dt}=i\langle [ H^r,\mathcal{A}(t)]\rangle +\langle \xi \mathcal{A}(t)\rangle,
\label{Eq5}
\end{equation}%
where the notation $\langle \cdots \rangle$ denotes an average on steady state and the Heisenberg operator $\mathcal{A}$ denotes the internal operator of the system. The superoperator $\xi $ describes the spontaneous emission of the internal excited state, that is
\begin{equation}
\xi \mathcal{A} =\sum_j\frac{\Gamma_{ji}}{2}(2 \sigma^{\dagger}_{ij}  \mathcal{A}\sigma_{ij}-\sigma_{jj}\mathcal{A}-\mathcal{A}\sigma_{jj}),
\label{Eq6}
\end{equation}
with $\sigma_{ij}=|i\rangle\langle j| $, $\Gamma_{ji}$ being the rates of decay from the excited state $|j\rangle $ down to the ground states $|i\rangle $.

On the other hand, dynamical evolution of the system is described by the master equation as
\begin{equation}
\dot{\rho}=-i[H^r,\rho]+\mathcal{D}(\rho,\Gamma),
\label{Eq12}
\end{equation}
where the superoperator $ \mathcal{D}(\rho,\Gamma)=\Gamma_{ji}(2 \sigma_{ij} \rho\sigma_{ij}^{\dagger}-\sigma_{jj}\rho-\rho\sigma_{jj})/2$ describes the spontaneous emission from the excited state $|j\rangle $ to the ground states $|i\rangle $. The steady state $\rho_{s}$ of the system is solved by making $\dot{\rho}_{s}=0$, and then the force in Eq. (\ref{Eq4}) can be numerically calculated. Numerical solution of friction coefficient can be obtained by the numerical derivative $dF/dv_z$ about the velocity $v_z$. As shown in Fig. \ref{SFig2}, the analytic result fits well the numerical simulation.

\section{solution of Tr$[\rho_s\sigma^{13}_y]$ and Tr$[\rho_s\sigma^{23}_y]$ for a three-level system}
The dynamics of a three-level system can also be described by Eq. (\ref{Eq5}), and in the superoperator $\xi $ there are two decay paths with the decay rate $\Gamma_{31}=\Gamma_1$ and $\Gamma_{32}=\Gamma_2$. In order to solve the Bloch equation, we rewrite $H_{2}^r$ in Eq. (\ref{Eq14}) as
\begin{equation}
H^r=\bar{\Delta} |1\rangle\langle 1|+\frac{\Omega}{2}\sigma_x^{13}+ \bar{\delta}|2\rangle\langle 2|+\frac{\bar{\Omega}}{2}\sigma_x^{23},
\label{Eq18}
\end{equation}
where the velocity-dependent detuning $\bar{\Delta}=(\Delta-k_1v_z\cos\theta)$ and $\bar{\delta}=(\delta+k_2v_z\cos\theta\cos\bar{\theta})$ with $z=v_z t$. Therefore, we obtain the Bloch equations of the operators as follows,
\begin{eqnarray}
\frac{d}{dt}\langle\rho_{11}\rangle&=&\frac{\Omega}{2}\langle\sigma_y^{13}\rangle+\Gamma_1\langle\rho_{33}\rangle, \notag \\
\frac{d}{dt}\langle\rho_{22}\rangle&=&\frac{\bar{\Omega}}{2}\langle\sigma_y^{23}\rangle+\Gamma_2\langle\rho_{33}\rangle, \notag \\
\frac{d}{dt}\langle\sigma_x^{12}&=&-(\bar{\Delta}-\bar{\delta})\langle\sigma_y^{12}\rangle+\frac{\Omega}{2}\langle\sigma_y^{23}\rangle+\frac{\bar{\Omega}}{2}\langle\sigma_y^{13}\rangle, \notag \\
\frac{d}{dt}\langle\sigma_y^{12}\rangle&=&(\bar{\Delta}-\bar{\delta})\langle\sigma_x^{12}\rangle+\frac{\Omega}{2}\langle\sigma_x^{23}\rangle-\frac{\bar{\Omega}}{2}\langle\sigma_x^{13}\rangle, \notag \\
\frac{d}{dt}\langle\sigma_x^{13}\rangle&=&-\bar{\Delta}\langle\sigma_y^{13}\rangle+\frac{\bar{\Omega}}{2}\langle\sigma_y^{12}\rangle-\frac{\Gamma_1+\Gamma_2}{2}\langle\sigma_x^{13}\rangle, \label{Eq19}  \\
\frac{d}{dt}\langle\sigma_y^{13}\rangle&=&\bar{\Delta}\langle\sigma_x^{13}\rangle+\Omega(\langle\rho_{33}\rangle-\langle\rho_{11}\rangle)-\frac{\bar{\Omega}}{2}\langle\sigma_x^{12}\rangle \notag \\
&&-\frac{\Gamma_1+\Gamma_2}{2}\langle\sigma_y^{13}\rangle, \notag \\
\frac{d}{dt}\langle\sigma_x^{23}\rangle&=&-\bar{\delta}\langle\sigma_y^{23}\rangle-\frac{\Omega}{2}\langle\sigma_y^{12}\rangle-\frac{\Gamma_1+\Gamma_2}{2}\langle\sigma_x^{23}\rangle, \notag \\
\frac{d}{dt}\langle\sigma_y^{23}\rangle&=&\bar{\delta}\langle\sigma_x^{23}\rangle+\bar{\Omega}(\langle\rho_{33}\rangle-\langle\rho_{11}\rangle)-\frac{\Omega}{2}\langle\sigma_x^{12}\rangle \notag \\
&&-\frac{\Gamma_1+\Gamma_2}{2}\langle\sigma_y^{23}\rangle, \notag
\end{eqnarray}
where we have $\langle\rho_{11}\rangle+\langle\rho_{22}\rangle+\langle\rho_{33}\rangle=1$. In general, the solutions for $\text{Tr}[\rho_s \sigma_y^{13}]$ and $\text{Tr}[\rho_s \sigma_y^{23}]$ are very complicated. Here we list them as
\begin{equation}
\langle\sigma_y^{13}\rangle=-\frac{\Omega\bar{\Omega}^2\Gamma_1M_1}{M_2+M_3+M_4},\ \langle\sigma_y^{23}\rangle=-\frac{\Omega^2\bar{\Omega}\Gamma_2 M_1}{M_2+M_3+M_4},
\label{Eq20}
\end{equation}
where $M_1=8(\bar{\Delta}-\bar{\delta})^2(\Gamma_1+\Gamma_2)$, $M_2=\bar{\Omega}^6\Gamma_1+\bar{\Omega}^4[2(4(\bar{\delta}-\bar{\Delta})\bar{\Delta}+\Omega^2)\Gamma_1+\Omega^2\Gamma_2]$, $M_3=\Omega^2\Gamma_2[(4\bar{\delta}(\bar{\Delta}-\bar{\delta})+\Omega^2)^2+4(\bar{\Delta}-\bar{\delta})^2(\Gamma_1+\Gamma_2)^2]$ and $M_4=\bar{\Omega}^2[4(\bar{\Delta}-\bar{\delta})^2\Gamma_1^3+2\Omega^2(4(\bar{\Delta}-\bar{\delta})^2+\Omega^2)\Gamma_2+8(\bar{\Delta}-\bar{\delta})^2\Gamma_1^2\Gamma_2+\Gamma_1(16(\bar{\Delta}-\bar{\delta})^2\bar{\Delta}^2+8(\bar{\Delta}-\bar{\delta})^2\Omega^2+\Omega^4+4(\bar{\Delta}-\bar{\delta})^2\Gamma_2^2)]$. Eq. (\ref{Eq20}) also shows $\sigma_y^{13}/\sigma_y^{23}=\bar{\Omega}\Gamma_1/\Omega\Gamma_2$.
On the other hand, we can also use Eq. (\ref{Eq12}) to numerically solve $\sigma_y^{13}$ and $\sigma_y^{23}$ by setting $ \mathcal{D}(\rho,\Gamma_1,\Gamma_2)=\Gamma_1 |1\rangle \langle 3 |\rho|3\rangle \langle 1|+\Gamma_2 |2\rangle \langle 3 |\rho|3\rangle \langle 2|-(\Gamma_1+\Gamma_2)(|3\rangle \langle 3|\rho+\rho|3\rangle \langle 3 |)/2$. The inset of Fig. \ref{SFig2} shows that the analytic curve fits perfectly the numerical curve.

Under the resonance condition $\bar{\Delta}=\bar{\delta}$, we can obtain
\begin{equation}
\langle\sigma_y^{13}\rangle=0, \quad \langle\sigma_y^{23}\rangle=0,
\label{Eq21}
\end{equation}
which means that there will be no cooling effect and this is called as dark resonance region. Moreover, since in our experiment $\Gamma_1\gg \Gamma_2$, we have $\langle\sigma_y^{13}\rangle\gg \langle\sigma_y^{23}\rangle$, and thus the cooling effect is mainly created by the first laser (see Fig. \ref{SFig4}). In the limit $\Gamma_2\rightarrow 0$, we obtain
\begin{equation}
\langle\sigma_y^{13}\rangle=-\frac{8(\bar{\Delta}-\bar{\delta})^2\Omega\Gamma_1}{\bar{N}},
\label{Eq22}
\end{equation}
with $\bar{N}=4(\bar{\Delta}-\bar{\delta})^2(4\bar{\Delta}^2+2\Omega^2+\Gamma_1^2)+8(\bar{\delta}-\bar{\Delta})\bar{\Delta}\bar{\Omega}^2+(\Omega^2+\bar{\Omega}^2)^2$. In the limit $\bar{\delta}\rightarrow 0$ and $\bar{\Omega}\rightarrow0$, we obtain $ \langle\sigma_y^{13}\rangle=-2\Omega\Gamma_1/[4\bar{\Delta}^2+2\Omega^2+\Gamma_1^2+\Omega^4/4\bar{\Delta}^2]$, which is different from Eq. (\ref{Eq9}). The reason is that in the limit $\bar{\delta}\rightarrow 0$ and $\bar{\Omega}\rightarrow0$, the above calculating method for three level system will diverge and leads to a wrong result.

\section{simulation for dynamical process of ions}
Motion of the trapped ions in the Doppler cooling process can be fully described by the Langevin equation. At a specific temperature $T$, the Langevin equation is given by
\begin{equation}
m\ddot{z}+\beta\dot{z}-\nabla\psi(z)=\xi(t),   \label{Eq38}
\end{equation}
where $m$ is the mass of a single ion, $z$ is the coordinate of the ion, $\psi(z)$ is the harmonic potential of the trap, i.e., $\psi(z)=m\omega_z^2z^2/2$ with the trap frequency $\omega_z$, $ \beta$ is the friction coefficient produced by the laser cooling and $\xi(t)$ is the stochastic force induced by temperature $T$ obeying the following ensemble average relations: $\langle \xi(t)\rangle=0$ and $\langle \xi(t)\xi(\tau)\rangle=2\beta k_BT\delta(t-\tau)$, where $k_B$ is the Boltzmann constant and $\delta$ denotes the Dirac $\delta$-function. In this equation, the friction coefficient, as an essential parameter, should be determined before the dynamical simulation is carried out. Our results, e.g., Eqs. (\ref{Eq11}) and (\ref{Eq24}), have just addressed this issue and provided analytical forms of $\beta$ for different energy-level cases. Besides, estimating ions' temperature based on our analytical results is very useful to optimize the Doppler cooling efficiency. For three-dimensional case, the ion's motion is described as
\begin{equation}
m\ddot{r}+\beta\dot{r}-\nabla\varphi(r)=\xi(t)
\label{Eq39}
\end{equation} 
where the position $r=(x,y,z)$ and $\varphi(r)$ is the three-dimensional trap potential.


\begin{thebibliography}{99}
\bibitem{note1} Here we assume that the first laser is in parallel with the $xz$ plane but the second laser beam, opposite to the first laser, is with a small angle $\bar{\theta}$ with respect to the $xz$ plane. This assumption is from the realistic situation in a recent experiment.
\end{thebibliography}
\end{document}